\documentclass[aps,prl,floatfix,twocolumn,superscriptaddress,showpacs]{revtex4-1}

\usepackage{amsmath,graphicx,amssymb,braket,xcolor,subfigure,upgreek}
\usepackage[colorlinks, linkcolor=blue, citecolor=blue, urlcolor=blue, breaklinks=true]{hyperref}

\pacs{71.35.-y,05.60.Gg,37.30.+i,81.05.Fb}

\begin{document}

\title{Cavity enhanced transport of excitons}
\author{Johannes Schachenmayer}
\affiliation{JILA, NIST, Department of Physics, University of Colorado, 440 UCB, Boulder, CO 80309, USA}

\author{Claudiu Genes}
\affiliation{Institut f\"ur Theoretische Physik, Universit\"at Innsbruck, Technikerstrasse 25, A-6020 Innsbruck, Austria}
\author{Edoardo Tignone}
\affiliation{IPCMS (UMR 7504) and ISIS (UMR 7006), Universit\'e de Strasbourg and CNRS, Strasbourg, France}
\author{Guido Pupillo}
\affiliation{IPCMS (UMR 7504) and ISIS (UMR 7006), Universit\'e de Strasbourg and CNRS, Strasbourg, France}

\date{\today}

\begin{abstract}
We show that exciton-type transport in certain materials can be dramatically modified by their inclusion in an optical cavity: the modification of the electromagnetic vacuum mode structure introduced by the cavity leads to transport via delocalized
polariton modes rather than through tunneling processes in the material itself. This can help overcome exponential suppression of transmission properties as a function of the system size in the case of disorder and other imperfections. We exemplify massive improvement of transmission for excitonic wave-packets through a cavity, as well as enhancement of steady-state exciton currents under incoherent pumping. These results may have implications for experiments of exciton transport in disordered organic materials. We propose that the basic phenomena can be observed in quantum simulators made of Rydberg atoms, cold molecules in optical lattices, as well as in experiments with trapped ions.
\end{abstract}

\maketitle

Understanding the transport properties of quanta and correlations and how to make this transport efficient over large distances are questions of fundamental importance in a variety of fields, ranging from experiments with cold atoms and ions \cite{jurcevic_quasiparticle_2014,richerme_non-local_2014,cheneau_light-cone-like_2012,gunter_observing_2013}, to quantum information theory \cite{bose_quantum_2007,lieb_finite_1972,nachtergaele_propagation_2006}, to (organic) semiconductor and solar cell physics \cite{forrest_path_2004, scholes_excitons_2006, menke_tailored_2013}.  
In most realistic situations, transport efficiency is known to be strongly inhibited by disorder. For example, Anderson-type localization  of single-particle eigenstates \cite{anderson_absence_1958} in disordered media implies an {\em exponential} suppression of  transmission, i.e.~over a distance of $N$ sites it decays as $T \propto \exp(-N)$. In this work we show how in general exponential suppression of energy transport via atomic and molecular excitons can be overcome by coupling the excitons to the structured vacuum field of a Fabry-Perot cavity placed transverse to the propagation-direction. 
In one dimension (1D), this trades the exponential suppression for a decay which is at most {\em algebraic}, $T\propto N^{-2}$, a massive enhancement that should be observable for realistic exciton-cavity couplings, system sizes, disorder strengths and even at room temperature \cite{orgiu_conductivity_2014,lidzey_strong_1998,lidzey_room_1999,holmes_strong_2004,coles_vibrationally_2011,schwartz_reversible_2011,kena-cohen_ultrastrongly_2013,plumhof_room-temperature_2014, balili_bose-einstein_2007, wertz_propagation_2012,alloing_evidence_2014}. 
While here we focus on {\em exciton}-transport, our work was originally inspired by first breakthrough experiments on {\em charge} transport in molecular semiconductors in the strong-coupling regime~\cite{orgiu_conductivity_2014}. In principle, the observed effect may open the way towards utilizing molecular materials as inexpensive and flexible alternatives to traditional silicon-based semi-conductors \cite{forrest_path_2004,shirakawa_synthesis_1977,tang_organic_1987,burroughes_light-emitting_1990,kim_micropatterning_2000,kahn_electronic_2003,sirringhaus_charge_2010,arias_materials_2010}.

\begin{figure}[t]
\includegraphics[width=0.9\columnwidth]{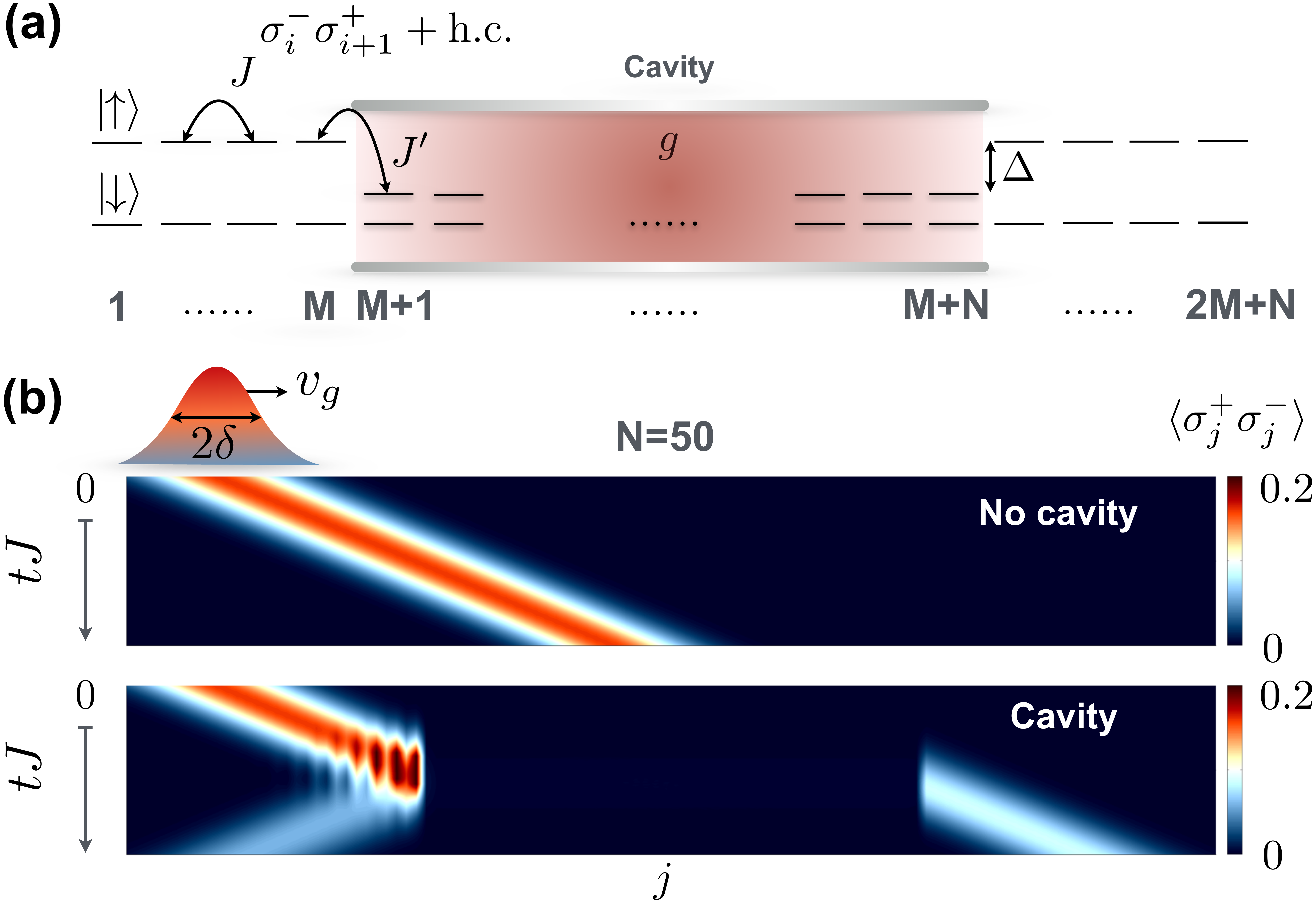}
\caption{\textit{Exciton transmission model}. Scheme of a chain of coupled two-level systems (tunneling rate $J$)
in which an exciton wave-packet propagates from the left into a cavity (group-velocity $v_g$) that is coupled to $N$ spins (cavity-spin coupling $g$). Under the right conditions, a large portion of the wave-packet can be almost instantaneously transmitted to the right side on a timescale $t \ll N/v_g$ [example in panel (b) with $N=50,v_g=2J, \delta=5, \Delta=69 J, J'=10J, g=10J$].} \label{fig1}
\end{figure}

Here, we provide a theoretical understanding of enhanced exciton transport for a model of two-level systems embedded in a cavity in the limit of strong collective light-exciton coupling.
We note that in these systems, strong collective coupling has been already demonstrated, and even used, e.g., to modify intrinsic material properties such as the work function \cite{hutchison_tuning_2013}.
On the other hand, our model also applies to artificial media such as cavity-embedded Rydberg lattice gases  \cite{low_experimental_2012,bettelli_exciton_2013}, polar molecules in optical lattices~\cite{yan_observation_2013,hazzard_many-body_2014}, or ions in linear Paul traps \cite{porras_effective_2004,friedenauer_simulating_2008}. In these systems, large couplings \cite{raimond_manipulating_2001} and reduced decoherence from spontaneous emission may allow for demonstrating essentially instantaneous coherent transport of excitonic wave-packets over  large distances with close-to-unit efficiency, $T\propto 1$. These experiments can analyze transport in systems with many excitations and in high dimensions, where modern numerical methods become inefficient \cite{vidal_efficient_2004,white_real-time_2004,daley_time-dependent_2004}.  This may contribute to improving our understanding  both of transport in real materials and of fundamental properties of information transport in strongly correlated light-coupled systems \cite{gong_persistence_2014,schachenmayer_entanglement_2013,jurcevic_quasiparticle_2014,richerme_non-local_2014}.

{\it The model} we consider consists of a chain of $N$ two-level systems or ``spins''  with local states $\ket{\uparrow}_i$ and $\ket{\downarrow}_i$  that are embedded in a cavity. The coupling to the cavity is governed by the Tavis-Cummings Hamiltonian $H_{\rm cav} = g \sum_{i}  \left( \sigma_i^+ a +\sigma_i^- a^\dag \right)$,
with $g$ the coupling strength, $a$ ($a^\dagger$) the destruction (creation) operator for the cavity photon, and $\sigma^\pm_i$ the Pauli spin raising/lowering operators for the spin at site $i$. We restrict our discussion to single excitations in the system. Such a localized excitation (i.e.~a state $\ket{\uparrow}_i$) has an energy $\omega_i$ ($\hbar \equiv 1$) and can tunnel between neighboring sites, as described by the Hamiltonian $H_0 = \sum_i \left[ \omega_i \sigma_i^+ \sigma_i^- -  J_i  \left( \sigma^+_i \sigma^-_{i+1} + \sigma^-_i \sigma^+_{i+1} \right) \right]
$. The tunneling rates $J_i$ can be site-dependent, and we define $J_i=J+ {\delta J_i}$, where $\delta J_i$ denotes random disorder drawn from a normal distribution with standard deviation $\delta J$. 
We note that in realistic physical setups the tunneling is typically induced by dipolar long-range forces that give rise to additional long-range hopping terms. These terms are not capable to lift the exponential suppression of transmission \cite{rodriguez_anderson_2003,SupMat}. 
Thus, for simplicity we consider the nearest neighbor tunneling model here. In addition,  a coupling between excitons and phonons can give rise to non-linear terms causing self-trapping effects \cite{rashba_critical_1994,agranovich_quantum_1999}. It can be shown that these terms are very small for realistic parameters and we will neglect them here.

The general dynamics of our system is governed by the master equation  $\dot \rho = -i [H,\rho] + \sum_\alpha \mathcal{L}_\alpha (\rho)$, with  $\rho$ the  density matrix and $H=H_0 + H_{\rm cav}$. The terms  $\mathcal{L}_\alpha (\rho) \equiv - \{L^\dag_\alpha L_\alpha, \rho\} + 2 L_\alpha \rho L_\alpha^\dag$ incorporate all  dissipative processes via ordinary Lindblad operators  $L_\alpha$. We consider cavity decay ($L_{\kappa} \equiv \sqrt{\kappa/2} a $) as well as  spontaneous emissions of each spin ($L_{{\rm sp.em.},i}\equiv \sqrt{\gamma_{\rm sp.em.}/2} \sigma^-_{i} $) or dephasing $(L_{{\rm deph.},i} \equiv \sqrt{\gamma_{\rm deph.}/2} \sigma^+_{i} \sigma^-_{i}$), deriving e.g.~from radiative decay and fluctuations in level-spacing (vibrations) due to the system being at finite temperature. In the homogeneous situation with $\omega_i=\omega_0$ and $J_i=0$, $H_{\rm cav}$ is responsible for the formation of dressed modes of the cavity photons and of the collective Dicke states $\sigma_0^\pm\equiv\sum_j \sigma_j^\pm/\sqrt{N}$, named as upper and lower polaritons for $u^\dagger\equiv(a^\dagger+\sigma_0^+)/\sqrt{2}$ and $d^\dagger\equiv(a^\dagger-\sigma_0^+)/\sqrt{2}$, respectively, with energy $\Omega_{u,d}=\omega_0\pm g\sqrt{N}$.  In this work we study two possible scenarios to observe  enhancement of exciton transport by exploiting these states: (i) a {\em wave-packet scattering} experiment, and (ii) steady state exciton currents under {\em incoherent pumping}. 


Case (i) is sketched in Fig.~\ref{fig1}(a): In addition to the $N$ spins in the cavity, $M$ spins are added to the left and  right of the cavity ($\mathcal{N}=2M+N$), coupled via $H_0$. We consider a homogeneous level-spacing inside and outside of the cavity, with $\omega_i=\omega_0 $ for $i=M+1,\dots,M+N$ and $\omega_i=\omega$ otherwise, and define  $\Delta=\omega-\omega_0$. We further denote  $J_i=J'$ for $i=M+1$ and $i=M+N$, i.e., at the entrance and exit of the cavity, to allow for impedance effects. At time $t=0$, a wave-packet of excitons, $|\psi(t=0)\rangle \propto \sum_{j=1}^{\mathcal{N}} e^{-i q_0 j}e^{-(j-j_0)^2/ (4 \delta^2)}|j\rangle$, with width $\delta$ (standard deviation) and initial quasi-momentum $q_0$ is injected on the left. Here, $\ket{j}\equiv \ket{\uparrow}_{j} \bigotimes_{i\neq j}
\ket{\downarrow}_i$ denotes the state of a single excitation at site $j$. The initial displacement from the cavity is $\delta_x=M-j_0$. As an example, we choose $\delta_x=20$, $\delta=5$ and $q_0=\pi/2$ [corresponding group velocity  $v_g=2J\sin(q_0)=2J$]. We are interested in the wave-packet fraction that for properly tuned parameters can be transferred nearly instantaneously to the right side of the cavity [cf.~Fig.~\ref{fig1}(b)]. 

Case (ii), in constrast, concerns a system with sites $i=1,\dots,N$ embedded in the cavity. Excitations are incoherently pumped to site $i=1$ from the left and removed from site $i=N$. This can be achieved via dissipative terms with $L_P \equiv \sqrt{\gamma_P/2} \sigma^+_{1}$ and $L_{\rm out} \equiv \sqrt{\gamma_{\rm out}/2} \sigma^-_{N}$, respectively. Under these conditions, we calculate the output exciton-current, $I_{\rm out}= {\rm tr} [n_e \mathcal{L}_{\rm out} (\rho)]$ (with {$n_e=\sigma^+_N \sigma^-_N$}) in the steady-state. Similar to the case \cite{manzano_quantum_2013}, this current arises naturally from the continuity equation ${d} \langle n_e \rangle/dt = 0 ={\rm tr} [n_e \mathcal{L}_P (\rho)] + {\rm tr} [n_e \mathcal{L}_{\rm out} (\rho)] + {\rm tr} [n_e \mathcal{L}_{\rm sp. em.} (\rho)] + {\rm tr} [n_e \mathcal{L}_{\rm deph.} (\rho)] -i {\rm tr}[n_e[H,\rho]]$. In the second part of this paper we show how $I_{\rm out}$ can be dramatically enhanced in the presence of the cavity.


{\it Wave-packet scattering --} In case (i), we first simplify the dynamics by neglecting dissipative terms and disorder (a valid approximation for e.g.~a Rydberg lattice gas \cite{SupMat}). Under these conditions, for $g=0$ the wave-packet ($v_g=2J$) reaches the right side of the cavity on a {\em long} timescale  $t_l J=\delta_x+2 \delta+ N/2$. This corresponds to the time required to hop over $N$ sites plus the time needed to enter and exit the cavity within the light-cone. Here we propose to use the polariton mode to tunnel $N$ sites almost instantaneously. 

The time-scale for a single excitation to couple in and out of such a mode is proportional to $\sqrt{N}/g$, and can be exceedingly small for large $g$. Then, transmission to the right side beyond the free-evolution light-cone is possible on an {\em ultra-short} scale $t_s J= \delta_x+2 \delta \ll t_l J$, limited only by the entrance time in the cavity. The dynamics can then be described via elastic scattering through the cavity, with a quasi-momentum dependent transmission function $T_q=|t_q|^2$, and $t_q$ the  coefficient appearing in the associated Lippmann-Schwinger equation \cite{ryndyk_green_2008}. 

The time-independent function $T_q$  determines the transmission properties of the material \cite{landauer_spatial_1957,buttiker_generalized_1985,biondi_self-protected_2014,longo_few-photon_2010}, and can be computed exactly for our model.  The coefficient has the general form $t_q=-2i\beta /[1+2i\beta]$, with $\beta=[2N J\sin(q)]^{-1}\sum_{n} |J'|^2/[\omega-2J\cos(q)-\tilde \Omega_n]$. Here, $\tilde \Omega_n$ is the $n$th eigenvalue of the reduced Hamiltonian for the cavity-coupled central $N$ sites of the chain \cite{SupMat}. The resulting $T_q$ in general presents three regions of ballistic transmission (i.e., $T_q=1$). These correspond to (a) ordinary exciton hopping for $\Delta \sim 0$, with an approximate width $4J$, as well as (b) two peaks for $\Delta \sim \Omega_{u,d}-J$. The latter correspond to polariton-mediated transmission, and have an approximate Lorentzian shape with a  $N$-dependent full width at half maximum (FWHM) $w = J'^2/(N |v_g|)$. For large enough strength of the collective exciton-cavity coupling $g\sqrt{N}> \max[w, 4J, \kappa]$ all peaks are well separated, which defines the {\it collective strong coupling} regime. In the following we focus on this regime, where in the vicinity of the polariton peaks $T_q$ is found to simplify to
\begin{align}
    T_{q} = \left\{1+N^2 J^2 \sin^2(q) [\omega +J(1-2\cos(q))-\Omega_{u,d}]^2/J'^4\right\}^{-1}.
    \label{eq:Tq}
\end{align}

Time-dependent wave-packet scattering can be investigated via numerical exact diagonalization. We define a time-dependent transmission as $T_{t'}$$=\sum_{j>M+N}^{} \langle \sigma^+_j \sigma^-_j \rangle_{t'}$,
which measures the total number of excitations that reach the right side of the system at a given time $t'$.

\begin{figure}[t]
\includegraphics[width=1\columnwidth]{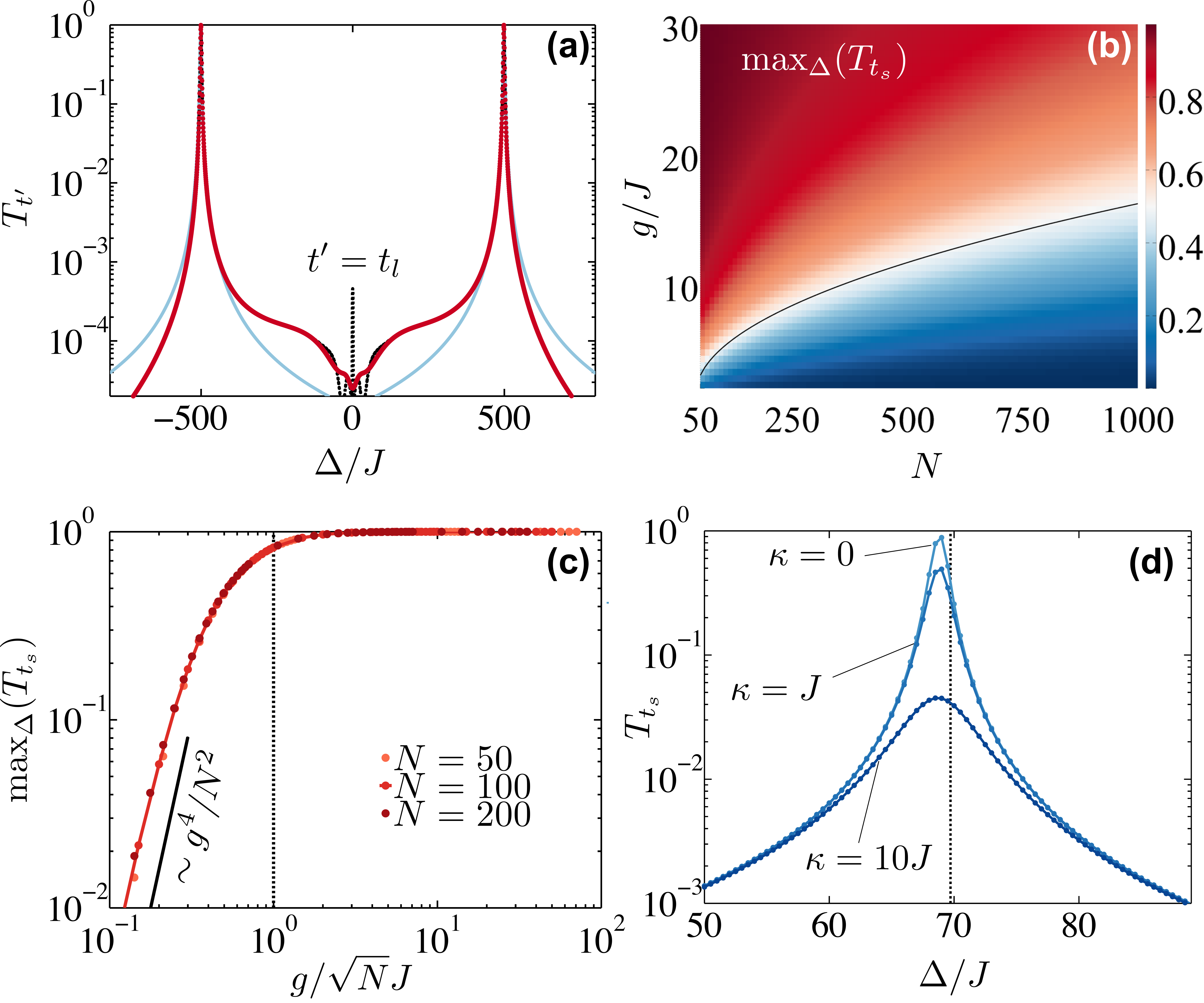}
\caption{\textit{Ultra-fast transmission} of a wave-packet with $\delta=5$, $\delta_x=20$, and $v_g=2J$. We choose  $J'=4 \tilde J_N$ (see text).  (a) Long-time and ultra-short transmission ($T_{t_l}$ and $T_{t_s}$) as function of $\Delta$. The cavity embeds $N=100$ sites and $g=50J$ (strong collective coupling regime). Clearly, two peaks of $T_{t_s}$ and $T_{t_l}$ appear at the polariton energies. The numerical calculation (red line) agrees with the analytical result (blue line).  $T_{t_l}$ (black dotted line) contains a small $\Delta \sim 0$ peak, corresponding to free evolution. (b)  $\max_\Delta (T_{t_s})$ as function of $g$ and $N$. To keep the ultra-fast transmission fixed, $g \propto \sqrt{N}$ (solid line) is required. (c) Crossover into the regime of large ultra-fast transmission around  $g\sim \sqrt{N} J$ (dotted line).  $\max_\Delta (T_{t_s})$ for $N=50,100,200$ is shown as function of $g/\sqrt{N}$ (on top of each other). For small $g$, $T_{t_s} \sim g^4/N^2$ (solid line). (d) Shrinkage and broadening of transmission peaks for finite cavity decay $\kappa$ ($N=M=50$, $g=10J$).}
\label{fig2}
\end{figure}

Our goal is to realize large ultra-fast transmission via the polariton peaks, i.e.~$T_{t'}\sim 1$ at $t'=t_s$. Two conditions have to be met: (i) The detuning $\Delta$ has to match the energy of one of the polariton peaks; and (ii) the wave-packet has to be sufficiently sharp in quasi-momentum space to fit into the energy window $w$, implying a real-space width on the order of the cavity length. While this can be generally difficult to realize, we find that condition (ii) can be satisfied by a choice $J'\propto \tilde J_N \equiv (2 \ln2)^{1/4} \sqrt{N/2\delta} J$ (ensuring an $N$ independent width), similar to an impedance effect.

In Fig.~2(a) we compare $T_{t'}$ for different $\Delta$, for  $t'=t_s$ (red continuous line) and $t'=t_l$  (black dashed line). We choose $N=100$, large $g=50J$, and set $J'=4 \tilde J_N$. As expected, we find the existence of two distinct polariton peaks, suitable for ballistic transmission on the ultra-fast scale  $t_s$. The position and width of the peaks are in agreement with the analytical time-independent predictions of Eq.~\eqref{eq:Tq}. The peak at $\Delta \sim 0$ instead reflectes regular exciton hopping on a time-scale $t_l \gg t_s$. Note that here $T_{t_l}<1$ due to backscattering at the cavity entrance where $J'>J$.

When decreasing the coupling strength $g$, the exciton dynamics through the cavity slows down considerably  \cite{SupMat}. The scattering becomes generally inelastic within $t_s$: part of the wave-packet energy remains in the cavity and $T_{t_s}<1$. However, we find that even for moderate couplings, a large fraction of the exciton wave-packet is transmitted within $t_s$.  This is shown in Fig.~\ref{fig2}(b). There, $\max_\Delta (T_{t_s})$ (i.e.~the best achievable $T_{t_s}$ for $\Delta$ chosen close to the upper polariton energy) is plotted as a function of $g$ and $N$: For increasing $N$, $T_{t_s}$ remains large and constant for a choice $g \sim \sqrt{N} J$. In addition, Fig.~\ref{fig2}(c) shows that $T_{t_s}$  vs. $g/\sqrt{N}J$ displays a universal behavior for different $N$. Here, $T_{t_s}$ reaches large values $\sim 80\%$ for $g=\sqrt{N} J$ improving to $100\%$ when increasing $g/\sqrt{N}J$. This is expected since for $g > \sqrt{N}J$, we enter the elastic scattering regime, in which the time-scale for coupling in and out of the polariton mode becomes negligible. Then, $T_{t_s}\sim 1$ is possible over arbitrarily large distances,  if a coupling strength $g \gtrsim \sqrt{N}J$ can be engineered. Interestingly, even for $g\ll \sqrt{N} J$ a significant part of the wave-packet is transmitted within $t_s$. In this regime (inelastic scattering and collective strong coupling), we find a general scaling of  $T_{t_s} \sim g^4$ and $T_{t_s} \sim 1/N^2$. Thus, cavity-mediated transmission decreases only {\it algebraically} with $N$, which can be important, e.g., when competing against exponential suppression due to disorder.

A lossy cavity ($\kappa \neq 0$) generally decreases $T_{t_s}$ because of loss of exciton population, while the FWHM increases accordingly. Fig.~\ref{fig2}(d) demonstrates that ultra-fast transmission of a large wave-packet fraction is still possible for  $\kappa\sim J$ (as, e.g.,~in a polar molecule setup).  We also note that for $\kappa \gg 1$ (after an adiabatic elimination of the cavity mode) dynamics can be described by an all-to-all Hamiltonian $ H_{\rm eff} \approx \frac{2 g^2}{\kappa} \sum_{i, j} \sigma_i^- \sigma_j^+$. Similar as with $ H_{\rm cav}$ above, we find that $H_{\rm eff}$ can give rise to ultra-fast transmission. We propose that such a situation could for example be observed in experiments with trapped ions, where these type of very long-ranged interactions arise naturally even in the absence of a cavity \cite{SupMat}.

\begin{figure}[t]
\includegraphics[width=1\columnwidth]{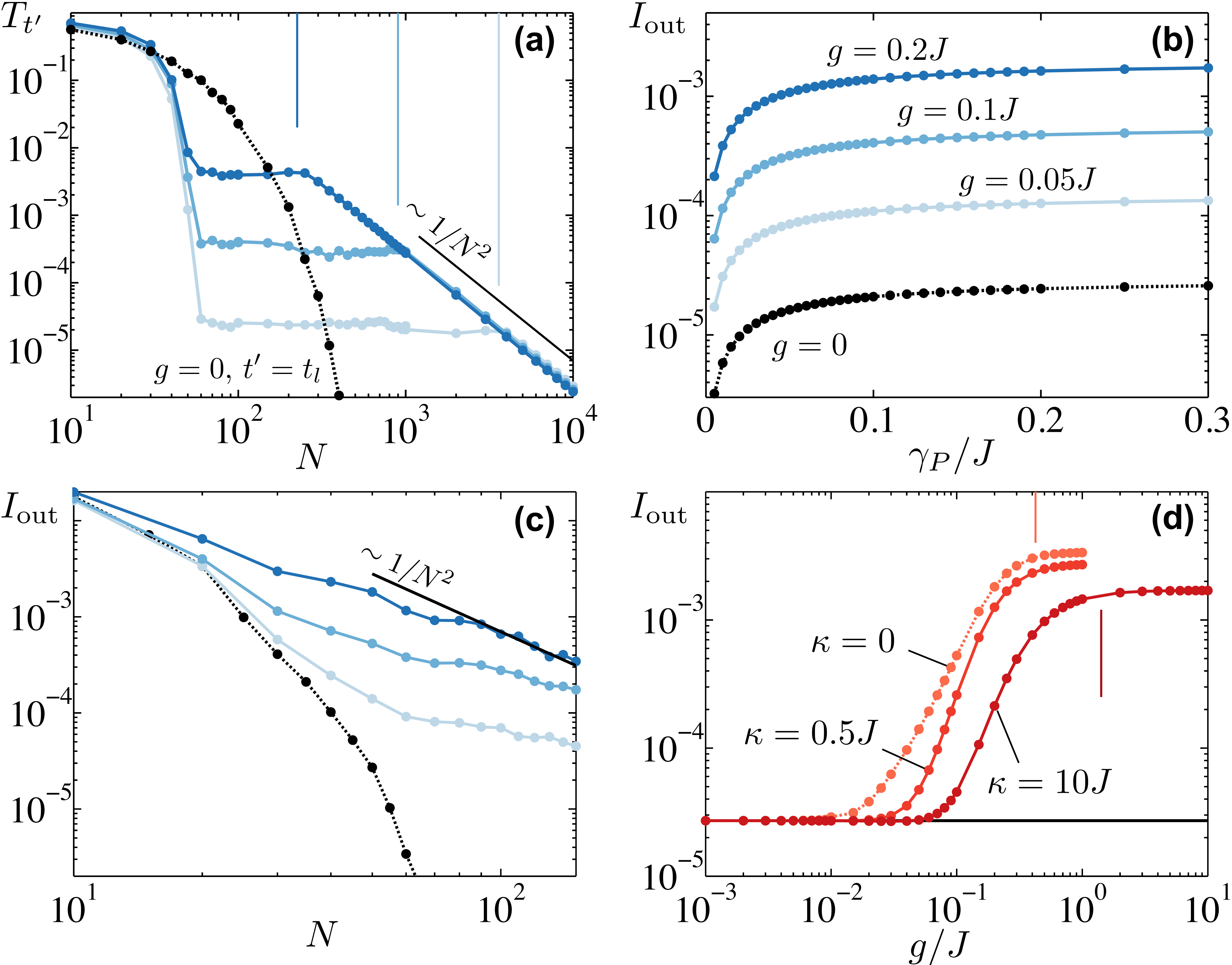}
\caption{\textit{Weakly coupled cavities.} Blue lines/poins denote  $g/J$ $=$ $0.05$, $0.1$, $0.2$ [light to dark]. $g=0$ is shown as black dashed line with points.  (a) $T_{t_s}$ and $T_{t_l}$ for wave-packet scattering [as in Fig.~2(a), $\Delta=g\sqrt{N}-J$] as function of $N$. We average over 200 disorder realizations in $J_i$ ($\delta_J=0.2J$). For $g=0$ we plot $T_{t_l}$, for $g>0$ $T_{t_s}$. $T_{t_l}$ decreases exponentially with $N$. Cavity-assisted $T_{t_s}$ becomes constant in the weak coupling regime and decays as $1/N^2$ [black solid line] in the collective strong coupling regime. Transitions between the regimes are indicated by vertical lines, marking $g\sqrt{N}=3J$. (b) Steady state exciton current under incoherent pumping, $I_{\rm out}$, for $N=50$ and in presence of disorder, spontaneous emissions, and dephasing (see text, $\gamma_{\rm out}=2J$). Small $g$ already leads to massive enhancement. (c) Dissipation and disorder leads to $I_{\rm out} \propto {\exp(-N)}$ for $g=0$ ($\gamma_P=0.5J$), while for $g>0$ $I_{\rm out}$ decays sub-exponential. In the collective strong coupling regime, $I_{\rm out}\sim 1/N^2$ [black solid line]. (d) $I_{\rm out}$ as function of $g$. The crossover to the collective strong coupling regime shifts for $\kappa>0$ ($N=50$, $\gamma_P=0.5J$). Vertical lines: $g\sqrt{N}=3J$ and $g\sqrt{N}=10J$, respectively. (b-d): $\gamma_{\rm sp. em.}=0.04J$, $\gamma_{\rm deph.}=0.9J$, and $\delta_J=0.2J$, single disorder realization.}\label{fig3}
\end{figure}

In realistic organic semi-conductors, disorder is key both in the spatial distribution and dipole orientation of molecules, implying site-dependent $J_i$ in $H_0$. In addition, typical cavity couplings are very small ($g\sim 0.1J$) \cite{SupMat}. Fig.~\ref{fig3}(a) shows $T_{t'}$ as function of $N$ for the same setup as in Fig.~\ref{fig2}, with $\delta J=0.2$ but for fixed $\Delta=g\sqrt{N}-J$. Without cavity ($g=0$) $T_{t_l}$ is exponentially suppressed, leading essentially to zero transmission for large $N$ ($T_{t_s}<10^{-6}$ for $N\gtrsim 400$ sites), as expected from Anderson-type localization \cite{anderson_absence_1958}. However, the localized eigenstates of the system can be modified by the cavity \cite{biella_dicke_2014,biella_subradiant_2013,celardo_interplay_2013}.  Adding weak cavity couplings ($g=0.05J, 0.1J,0.2J$) already lifts the transport suppression and allows for a small but finite transmission even for systems with $N=10^4$ sites. Consistent with the discussion above, in the collective strong coupling regime [right of vertical lines] we find an universal algebraic behavior $T_{t'} \sim 1/N^2$. Interestingly, even in the weak coupling regime  [left of vertical lines], i.e.~when the two polariton peaks are not resolved, we find a small constant transmission orders of magnitudes above the cavity-free case.

We note that in this paper we deal with a 1D situation, but we expect the main findings to also hold for dimensionalities $d>1$. While relative improvement of $T_{t_l}$ compared to the $g=0$ case can decrease with increasing $d$ (excitations can tunnel past impurities more easily), a finite $T_{t_s}$ for $g=0$ will also be impossible for $d=2,3$ because of the Lieb-Robinson bound \cite{lieb_finite_1972}. In contrast the cavity-photon mode occurs in any dimensions and thus one can expect the transmission mechanism to work in arbitrary dimensions. Since Anderson localization is also present in 2D, and in 3D below the mobility edge \cite{abrahams_scaling_1979}, an exponential improvement of transmission can be expected in such a situation

{\it Incoherent pumping setup --} We now consider the case (ii), i.e.~we analyze steady-state currents $I_{\rm out}$ that develop under incoherent pumping of excitations ($\gamma_P,\gamma_{\rm out}>0$). Spontaneous emission and dephasing are now included with  $\gamma_{\rm sp. em.}=0.04J$ and $\gamma_{\rm deph.}=0.9J$, respectively. Pump rates $\gamma_P$ play the role similar to a ``voltage'' but for exciton currents. We plot $I_{\rm out}$ - $\gamma_P$ curves in Fig.~\ref{fig3}(b). The figure shows that even small cavity couplings $g$ can increase $I_{\rm out}$ by orders of magnitude compared to the cavity-free case. This finding is in stark contrast with previous works with exciton-polaritons in multi-mode cavities ~\cite{agranovich_nature_2007} and constitutes one of the key results of this work.

Consistent with the wave-packet dynamics above, Fig.~\ref{fig3}(c) shows that, for $g=0$, $I_{\rm out}$ decreases exponentially with $N$, due to the various dissipative terms and the disorder. However, choosing $g=0.05,0.1,0.2$, changes the currents dramatically: for $N=150$ and $g=0.2$ the collective strong coupling regime is barely reached $g\sqrt{N}\sim 2.5 J$; nevertheless, remarkably, we find that $I_{\rm out}$, just as $T_{t_s}$ above, already displays an algebraic $1/N^2$ decrease. The fact that the cavity enhancement of $I_{\rm out}$ is induced by a collective cavity coupling is further demonstrated in Fig.~\ref{fig3}(d), where $I_{\rm out}$ is shown vs. $g$, for a few values of $\kappa$. A sudden increase of $I_{\rm out}$ occurs when $g$ exceeds a particular value (vertical lines). By inspection, we find that this indeed corresponds to the point where $g\sqrt{N}$ exceeds all other energy scales. Consistently, this point is shifted to larger values of $g$ for large $\kappa=10$.

{\it Conclusion \& Outlook --} In this work, we have shown that both incoherent and coherent exciton transport in a spin chain can be dramatically enhanced by collective coupling to the structured vacuum field of a Fabry-Perot cavity. These results may be relevant for disordered organic semiconductors  at room-temperature, where exciton conduction may be ameliorated by orders of magnitude, as well as for artificial media made of Rydberg atoms, polar molecules or cold ions at sub-mK temperatures. It is an exciting prospect to investigate whether strong coupling can also induce the ultra-fast propagation of classical and quantum correlations \cite{schachenmayer_entanglement_2013,gong_persistence_2014,litinskaya_exciton_2008}. 
While in 1D a modified density matrix renormalization group technique \cite{vidal_efficient_2004,white_real-time_2004,daley_time-dependent_2004} might provide us with an answer, the higher dimensional situation could be a first example where only the artificially engineered quantum simulator setups can do so. Finally, a key open challenge not addressed here is to explore the physical mechanisms behind the enhancement of charge conductivity as reported in experiments \cite{orgiu_conductivity_2014}.

{\it We note} that results for transport in disordered organic semiconductors related to those reported here have been independently obtained by J. Feist and F. J. Garcia-Vidal  \cite{feist_2014}.

{\it Acknowledgements--} We thank T.~W.~Ebbesen, F.~J.~Garcia Vidal, J.~Feist, L.~M.~Moreno and H.~Ritsch for fruitful discussion. J.S. acknowledges hospitality from the Institute for Theoretical Physics at the University of Innsbruck and the University of Strasbourg. This work was supported by the ERC-St Grant ColdSIM
(No. 307688), EOARD, and UdS via Labex NIE and IdEX, the JQI, FWF-ANR via BLUESHIELD, and RYSQ, the NSF PFC at the JQI and JILA (PIF-1211914 and PFC-1125844), as well as the Austrian Science Fund (FWF) via project P24968-N27 (CG). Computations utilized the HPC UdS, and the Janus supercomputer, supported by NSF (CNS-0821794), NCAR and CU Boulder.

\bibliography{exciton_cavity_transport}

\begin{widetext}

\pagebreak

\section*{SUPPLEMENTAL MATERIAL: CAVITY ENHANCED TRANSPORT OF EXCITONS}

\subsubsection{Details on scattering calculation}

It is convenient to write the reduced Hamiltonian for the cavity-coupled central $N$ sites in its eigenbasis, $\sum_n \tilde \Omega_n\Pi_n^{\dagger}\Pi_n$, where the projectors $\Pi_n=\ket{\rm vac}\bra{n}$ destroy an excitation in the eigenstates $\ket{n}$. Since the total Hamiltonian $H=H_0+H_{\rm cav}$ commutes with the excitation operator, i.e. $\sum_{i=1}^{\mathcal{N}}\sigma_i^{\dagger} \sigma_i^-+\sum_n\Pi_n^{\dagger}\Pi_n$, the one excitation ansatz can be written as
\begin{equation}
\ket{\psi_q}=\left[\sum_{i=1}^MC^{(l)}_{i,q}\sigma_{i}^{\dagger}+\sum_n p_q^n P_n^{\dagger}+\!\!\sum_{i=M+N+1}^{\mathcal{N}}C^{(r)}_{i,q}\sigma_{i}^{\dagger}\right]\ket{\rm vac}.
\end{equation}
For the single exciton scattering $C^{(l)}_{i,q}\approx e^{-iq(i-M)}+r_q e^{iq(i-M)}$ and $C^{(r)}_{i,q}\approx t_qe^{iq(i-M-N-1)}$. The Schr\"odinger equation $H\ket{\psi_q}=\omega_q\ket{\psi_q}$ with $\omega_q=\omega-2J\cos(q)$ entails $t_q=-2i\beta /(\Gamma_l\Gamma_r+|\beta^2|)$  with:
\begin{equation}
\Gamma_{l,r}=1+\frac{i}{2v_g}\sum_n\frac{|J'_{n,l,r}|^2}{\omega_q-\tilde\Omega_n} \quad \beta=\frac{1}{2v_g}\sum_n\frac{J'_{n,l}J'^*_{n,r}}{\omega_q-\tilde\Omega_n}.
\end{equation}
Couplings $J'_{n,l}=-J'\bra{n}\sigma_{M+1}^{\dagger}\ket{\rm vac}$ and $J'_{n,r}=-J'\bra{n}\sigma_{M+N}^{\dagger}\ket{\rm vac}$ involve exciton amplitudes at the leftmost and rightmost cavity-coupled sites.
The $N+1$ eigenvalues $\tilde \Omega_n$ are found by solving the reduced Schr\"odinger equation for the central $N$ sites. They obey the non-linear equation
\begin{equation}
\label{eq:nl}
\frac{g^2}{N(\tilde \Omega_n-\omega_0)}\sum_{k=1}^{N}\frac{\left(\sum_{i=1}^N\alpha_k^i\right)^2}{\tilde \Omega_n -\omega_k}=1,
\end{equation}
with $\alpha_k^i=\sqrt{2/N+1}\sin(\pi/(N+1)kj)$ satisfying $\sum_{k=1}^N\alpha_k^i\alpha_k^{i'}=\sum_{k=1}^N\alpha_i^k\alpha_{i'}^{k}=\delta_{i,i'}$.

However, open boundary conditions (OBC) do not allow for an analytical expression of the transmission amplitude $t_q$; it can be obtained by using periodic boundary conditions (PBC). Fourier transforms of the spin operators,
\begin{equation}
\sigma_k^{\dagger}=\frac{1}{\sqrt{N}}\sum_{j=M+1}^{M+N}\sigma_j^{\dagger}e^{i\frac{2\pi}{N} kj},
\end{equation}
allow to factorize the superradiant mode $\sigma_0^{\pm}$ as well as to introduce the polaritons through the transformations $u^{\dagger}=(a^{\dagger}+\sigma^{\dagger}_0)/\sqrt{2}$ and $d^{\dagger}=(a^{\dagger}-\sigma^{\dagger}_0)/\sqrt{2}$. The Hamiltonian can then be diagonalized as $\tilde \Omega_u \ket{u}\bra{u}+\tilde \Omega_d \ket{d}\bra{d}+\sum_{k=1}^{N-1}\tilde \Omega_k\sigma_k^{\dagger}\sigma_k^-$ with polariton energies $\tilde \Omega_{u,d}=\omega_0-J\pm g\sqrt{N}$. The energies of the $N-1$ uncoupled cavity modes are instead $\tilde \Omega_k=\omega_0-2J\cos(2\pi k/N)$. These also make the evaluation of coefficients $\beta$ and $\Gamma_{l}=\Gamma_{r}=\Gamma$ straightforward:
\begin{equation}
\beta=\frac{|J'|^2}{2NJ\sin(q)}\left[\frac{1}{\omega_q-\tilde\Omega_u}+\frac{1}{\omega_q-\tilde\Omega_d}+\sum_{k=1}^{N-1}\frac{1}{\omega_q-\tilde\Omega_k}\right],
\end{equation}
\begin{equation}
\Gamma=1+i\beta.
\end{equation}
In fact $T_q=1/(1+\beta^{-2}/4)$. We also looked for transmission resonances in case of a site-dependent coupling $g_i$. In this case Eq.~(\ref{eq:nl}) reduces to
\begin{equation}
\frac{1}{N(\tilde \Omega_n-\omega_0)}\sum_{k=1}^{N}\frac{\left(\sum_{i=1}^N g_i \alpha_k^i\right)^2}{\tilde \Omega_n -\omega_k}=1,
\end{equation}
and the polariton peaks are at $\tilde\Omega_{u,d}=\omega_0-J\pm\sqrt{\sum_{i=1}^N g_i}$.

\begin{figure}[t]
\includegraphics[width=0.9\textwidth]{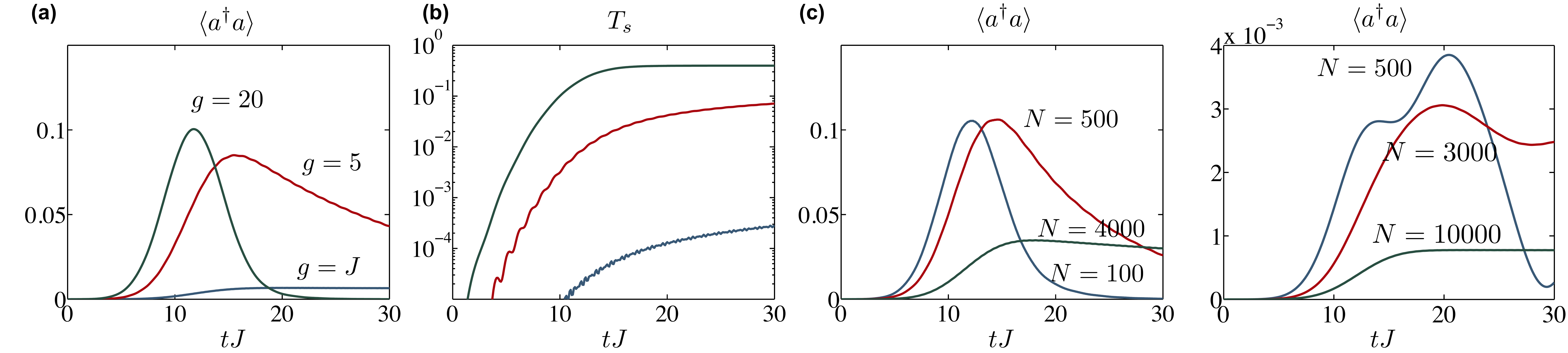}
\caption{\textit{Time-evolution in different regimes --}  (a) Evolution of the cavity occupation for the parameters of Fig.~2. $N=1000$. The crossover from the $\sim g^4$ to the $\sim g$ regime happens for $g\gtrsim 10J$. This is exactly the value below which excitations remain in the cavity and the scattering theory breaks down. (b) The corresponding time-evolution of the transmission. Only for $g=20$ the value saturates on the short time-scale. (c) The same crossover is visible when keeping $g=5J$ fixed and increasing $N$, i.e. this crossover happens when the $\sim 1/\sqrt{N}$ scaling changes to a $\sim 1/N^2$ scaling. (d) In the case of disorder we always have the situation that excitations remain in the cavity. Interestingly for systems up to $N=3000$ complex dynamics of the cavity occupation takes place. For larger $N$ we have a similar situation as in the case without disorder and small $g$, i.e. ~part of the excitation is stuck in the cavity (and constant). Note also that in the disorder case the occupation is much smaller.} \label{fig:scaling}
\end{figure}

\subsubsection{Comparison analytics and numerics}

We check under which circumstances our scattering theory gives agreement with the numerics. The  condition that has to be met is that we have to be in the elastic scattering limit, i.e. the excitation goes fully  in and out of the cavity on the time-scale of the experiment/simulation. Whether this is met can be checked by looking at the time evolution of the occupation number of the cavity mode.  We analyze this in Fig.~\ref{fig:scaling}, where we show examples of the different regimes.
In addition we check to what extend the height and peak-position of our transmission spectrum, which we calculate fully numericaly, agrees with the analytical expectation. Results are shown in Fig.~\ref{fig:strong_coupling}

\begin{figure}[b]
\includegraphics[width=0.9\textwidth]{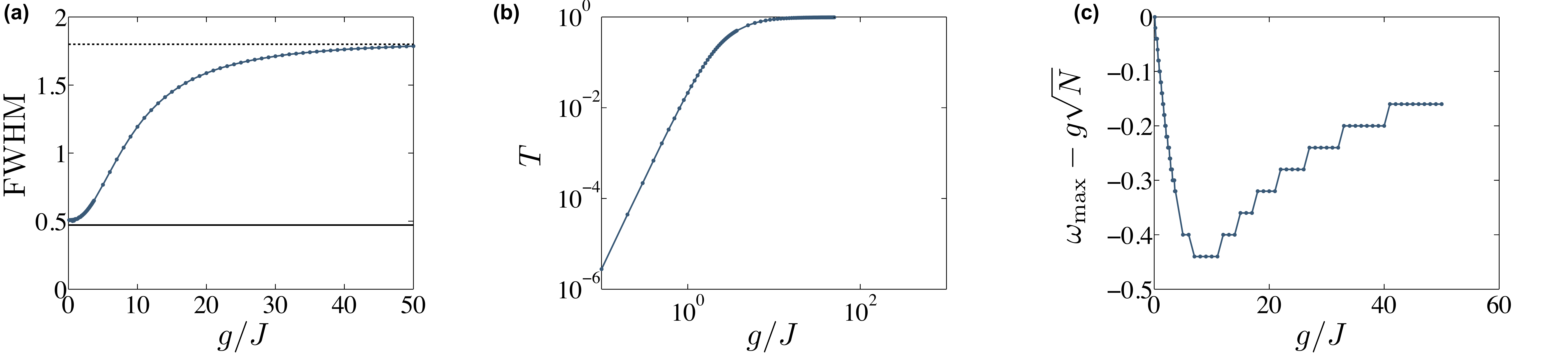}
\caption{\textit{Numerical evaluation of the transmission peaks}  (a) The FWHM  as function of $g$ for the numerical results from the scattering experiment ($N=100$, $\delta_x=20$, $\delta_0=5$, $J=0$, $J'=1.5\sqrt{2N/\delta}$. The lower solid line indicates the expected energy width of the wave packet. The upper dashed line is the expected transmission peak width from our analytical calculation(b) The corresponding transmission at the maximum peak position. The second kink from Fig.~2 disappears (and is thus related to the detuning) (c) the detuning from the analytical expected polariton peak-position.} \label{fig:strong_coupling}
\end{figure}

\subsubsection{Organic materials}

The dipole moment for typical molecules is $d=e \times0.75\,\text{nm}$ . Given a spacing of $x=3\,\text{nm}$ this yields a tunneling constant between nearest neighbors of $J=d^2/(4 \pi \epsilon_0 x^3)\approx 0.03\,\text{eV}$. According to \cite{orgiu_conductivity_2014} a Rabi-splitting of $\Omega_R \approx 1\,\text{eV} = 2 g \sqrt{N}$ can be achieved for $10^5$ molecules. Thus, a value of $g\approx0.0016 \,\text{eV}\approx0.05 J$ is realistic. Typical noise in the position is given by $\delta x=0.2\,\text{nm}$, which is a fluctuation of $7\%$ in $x$ and yields a fluctuation of $20\%$ in $J$.  Finally, a typical level spacing is $\omega=2\,\text{eV}\approx 70 J$.

\subsubsection{Atomic, molecular and optical systems}

\begin{figure}[t]
\includegraphics[width=0.35\textwidth]{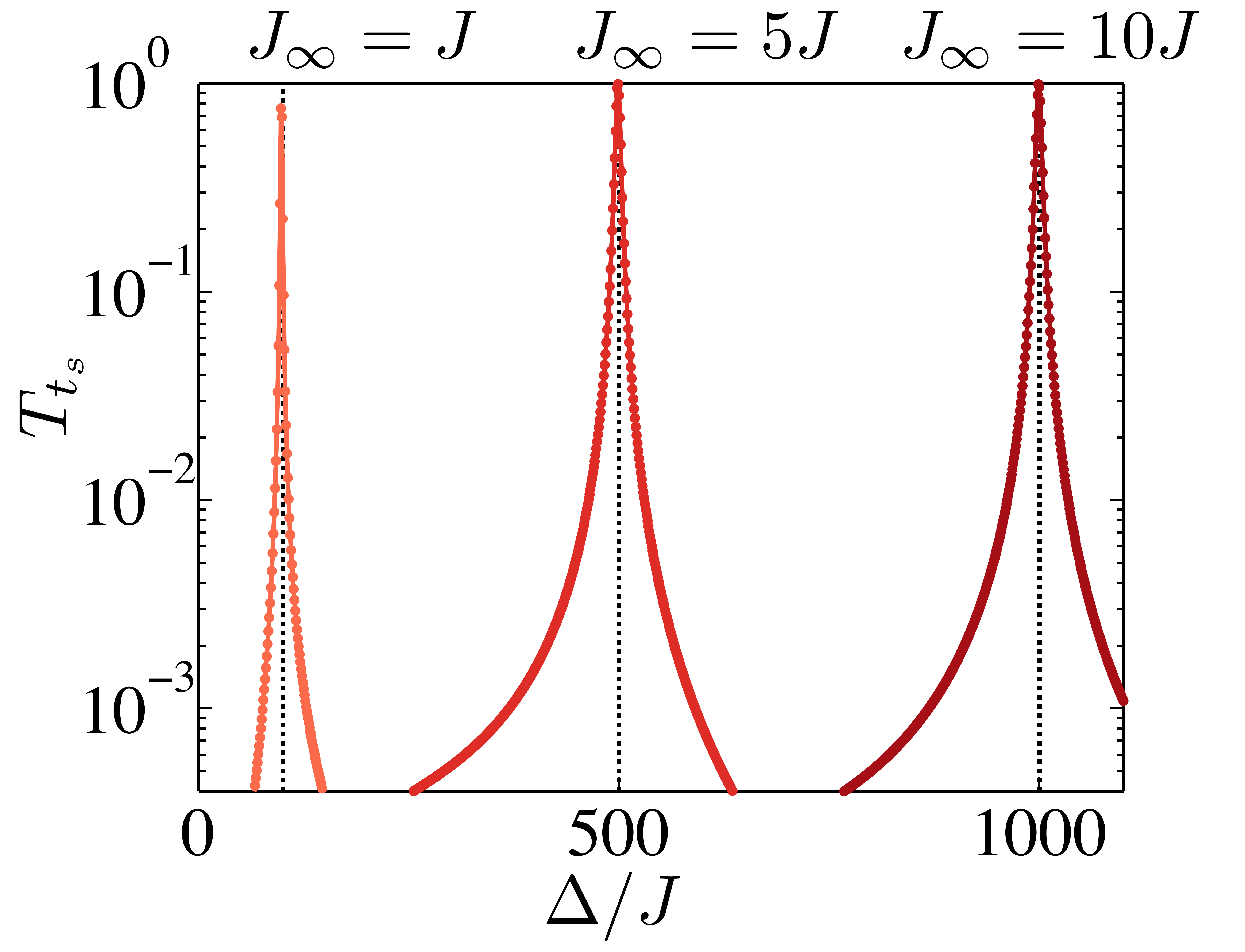}
\caption{\textit{All-to-all interactions --}  Wave-packet identical to the one in Fig.~2, $N=100$. The $N$ particles are coupled by $H_{\infty}$ (no cavity). $T_{t_s}$ as function of $\Delta$ shows peak at the values of the eigenenergy of the dominant eigenvalue of $H_{\infty }$\label{fig:a2a} }
\end{figure}

{\it Rydberg atoms -- } The XY model we consider in this manuscript can be realized in a {\em Rydberg lattice gas} \cite{bettelli_exciton_2013}. Here with a first pulse a Rydberg $\ket{n S}$ state lattice is created by using the dipole blockade. Then these states are coupled to another $\ket{n'P}$ state that serves as second spin state. For example, in \cite{bettelli_exciton_2013} the states $\ket{60S_{1/2}}$ and $\ket{59 P_{3/2}}$ are used ($^{87}$Rb atoms). The level spacing is $18.5\,\text{GHz}$. A transition dipole moment between these states is Rydberg-typically ($d \sim n^2$) very large and on the order of $d=2000\,ea_0$. Resonant microwave cavities for transition frequencies of $51\,\text{GHz}$ with dipole moments of $1000\,ea_0$ and coupling strengths of $g \approx 300\,\text{kHz}$ (Q-factor of $3 \times 10^8$) have been successfully engineered \cite{raimond_manipulating_2001}. Since $g \sim d  \sqrt{\omega_0}$, couplings of $g\approx 350\,\text{kHz}$ are within reach. On the other hand, for a separation of $20\,\mu\text{m}$, typical nearest neighbor tunneling is on order of $80 \,\text {kHz}$ \cite{robicheaux_simulation_2004} and thus $g$ much larger than $J$ is clearly possible. For a Q factor of $10^8$, the decay rate would be  $1\,\text{kHz}$ and thus negligible compared to $g$ and $J$. The lifetime of the Rydberg states can be on the order of tens of milliseconds and therefore also spontaneous emissions can be safely neglected.\\

{\it Polar molecules -- } The same type of  microwave cavities could be used for systems of {\em polar molecules} in optical lattices, where rotational states (spacing typically $\sim 2\,\text{GHz}$) can be used as spin-states. In these systems $J \approx 50\,\text{Hz}$ has been successfully observed in a recent experiment \cite{yan_observation_2013,hazzard_many-body_2014}. While much stronger couplings $g\gg J$ can be engineered for these systems, a challenge might be to build a  cavity with a sufficiently large Q-factor. For example, for $Q=10^8$, $\kappa=125\,\text{Hz}$ and thus larger than $J$. The lifetime of the states is sufficiently long to observe coherent dynamics over $\sim 0.1\,\text{s}$ \cite{yan_observation_2013}.\\

{\it Cold ions -- } In the domain of optical transitions, setups with {\em ions} in linear Paul traps might be considered. In these experiments tunneling rates of $J\approx 400,\text{Hz}$ can be achieved \cite{richerme_non-local_2014}. Note that since in these experiments the hopping is mediated by motional degrees of freedom of the ion-crystal, also long-range hoppings are important. Typical decay exponents range from $\alpha=0.1$ (almost all-to-all interactions) to $\alpha=2$. In the all-to-all case no cavity is required at all. Nevertheless, ion in  cavities with couplings of $g \sim 10\,\text{MHz}$ (with $\kappa \sim \text{MHz}$) \cite{stute_tunable_2012} can be engineered. Thus, here also very strong coupling in the regime $\kappa \gg J$ could be achieved.  We also numerically verified ultra-fast transmission in the case of an all-to-all coupling $H_{\infty}=J_\infty \sum_{i, j} \sigma_i^- \sigma_j^+$.  In Fig.~\ref{fig:a2a} we use the same wave-packet as in Fig.~2 of the main manuscript. Instead of coupling the sites $i=M+1,\dots,M+N$ to a cavity however, we couple them collectively via $H_\infty$. Again we find ultra-fast transmission peaks with $T_{t_s}=1$, however now they appear at values of $\Delta\sim\Omega_0= N J_\infty$, which corresponds to the eigenenergy of the dominant eigenvalue of the all-to-all Hamiltonian.

\subsubsection{Dipole-dipole interactions}

The model we consider in our manuscript is a toy model that has applications to the above mentioned systems. With exception of the ion trap situation, spin-exchange interactions are usually mediated via dipole-dipole interactions. Here we show that in most situations our simple nearest-neighbor hopping model is a good approximation to real dipole-dipole interactions.

Dipole-dipole interactions can give rise to elastic and inelastic interactions and also include retardation effects. As long as $ka \ll 1$, where $k$ is the wave-number of the dipole transition and $a$ is the average spacing between the dipoles, retardation effects as well as inelastic interactions can be neglected~\cite{lehmberg_radiation_1970}. In such a situation the interactions reduces to the usual dipole-form: \\\
\begin{align}
H_{dd}=\sum_{i,j; i\neq j} \frac{\bar J}{|i-j|^3} \left( \sigma^+_i \sigma^-_j +   \sigma^+_j \sigma^-_i  \right)
\label{ddham}
\end{align}
For our different dipole systems (taking the numbers from the previous section and the corresponding references), one finds $ka\sim 10^{-2}\pi$ (organic molecules), $ka\sim10^{-8}\pi$ (Rydberg atoms), $ka\sim10^{-7} \pi$ (Polar molecules). Thus, we are certainly in the regime where Eq.~\eqref{ddham} is valid.

In the main manuscript we restricted interactions to nearest-neighbors. To test whether the long-range interactions have any significant effect we perform calculations with and without these additional extra terms. We use parameters equivalent to those in Fig. 3a of the main manuscript. In Fig.~\ref{figA1}, a wave-packet is propagating into an area where the dipoles are randomly displaced by $5\%$ of the lattice constant (average over $50$ disorder realization). We plot the long-time transmission $T_{t_l}$. In  one case we only keep the nearest neighbor tunneling terms, whereas in the other case we keep the full dipole interactions $\propto |i-j|^{-3}$ between all spins. While long-range hopping terms can improve the transmission, we find that they can not turn the exponential decrease into an algebraic one. Adding however a weakly coupled cavity ($g=0.2J$)  clearly turns the exponential decay into an algebraic one.

\begin{figure}[tb]
\centering
\includegraphics[width=0.3\textwidth]{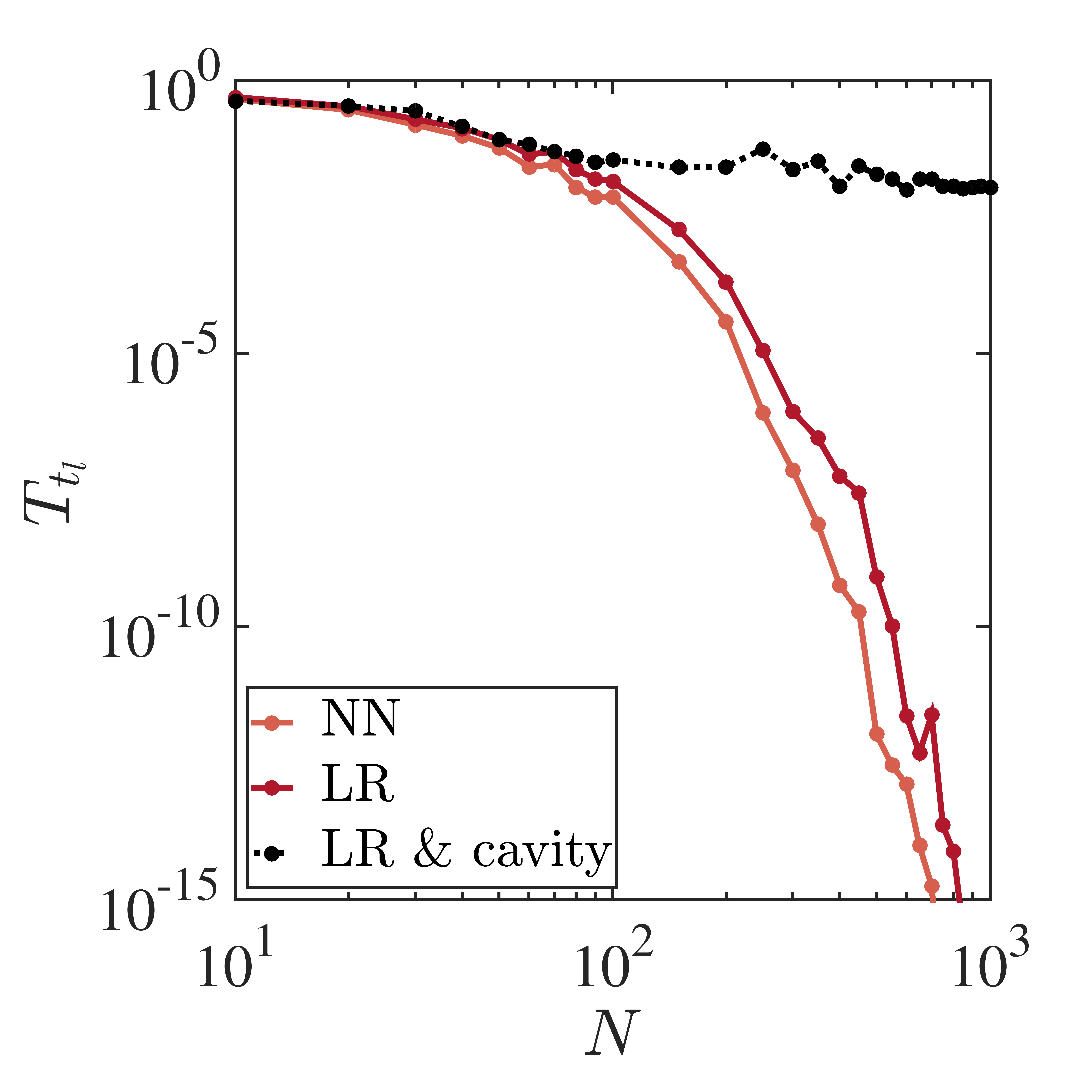}
\caption{Long-time wave-packet transmission as function of the system size $N$ (identical parameters as in Fig.~3a in the main manuscript). We show $T_{t_l}$ for $50$ disorder realizations with a displacement standard deviation of $5\%$ of the lattice constant. Shown are transmissions with nearest-neighbor tunneling terms only (NN) and with all dipolar long-range hopping terms (LR). The LR terms increase transmission but cannot lift the exponential suppression. Only in the case of a cavity ($g=0.02J$) the transmission scales algebraically with  $N$.  \label{figA1}}
\end{figure}

\end{widetext}


\end{document}